\documentclass[twocolumn,showkeys,superscriptaddress,preprintnumbers,amsmath,amssymb,showpacs,prl]{revtex4-1}
\usepackage{graphicx}
\usepackage{dcolumn}
\usepackage{bm}
\usepackage{amsmath}
\usepackage{xcolor}

\begin{document}
	
\title{Breakdown of Scaling and Friction Weakening in the Critical Granular Flow}

\author{Andrea Baldassarri}
\affiliation{CNR - Istituto dei Sistemi Complessi, Dipartimento di Fisica, Sapienza Universit\`a,  P.le A. Moro 2, 00185 Roma - Italy}
\author{Mario A. Annunziata}
\author{Andrea Gnoli}
\affiliation{CNR - Istituto dei Sistemi Complessi, Dipartimento di Fisica, Sapienza Universit\`a,  P.le A. Moro 2, 00185 Roma - Italy}
\author{Giorgio Pontuale}\email[]
{Presently at Consiglio per la Ricerca in Agricoltura e l'Analisi dell'Economia Agraria (CREA), Via Santa Margherita 80, I-52100 Arezzo, Italy}
\affiliation{CNR - Istituto dei Sistemi Complessi, Dipartimento di Fisica, Sapienza Universit\`a,  P.le A. Moro 2, 00185 Roma - Italy}
\author{Alberto Petri}
\affiliation{CNR - Istituto dei Sistemi Complessi, Dipartimento di Fisica, Sapienza Universit\`a,  P.le A. Moro 2, 00185 Roma - Italy}

\begin{abstract}
The way granular materials response to an applied shear stress is of the
utmost relevance to both human activities and natural environment. One
of the their most intriguing and less understood behavior, is the
stick-instability, whose most dramatic manifestation are earthquakes,
ultimately governed by the dynamics of rocks and debris jammed within
the fault gauge. Many of the features of earthquakes, i.e.
intermittency, broad times and energy scale involved, are mimicked 
by a very simple experimental set-up, where small beads of glass under
load are slowly sheared by an elastic medium. Analyzing data from long
lasting experiments, we identify a critical dynamical regime, that can
be related to known theoretical models used for "crackling-noise"
phenomena. In particular, we focus on the average shape of the slip
velocity,  observing a "breakdown of scaling": while small slips show a
self-similar shape, large does not, in a way that suggests the presence of subtle inertial effects within the granular system. In order to characterise the crossover between the two regimes, we investigate the frictional response of the system, which we trat as a stochastic quantity. Computing different averages, we evidence a weakening effect, whose Stribeck threshold velocity can be related to the aforementioned breaking of scaling.

\end{abstract}

\keywords{granular matter, friction, avalanche, crackling dynamics, scaling}

\pacs{45.70.-n 45.70.Ht 05.65.+b 07.79.Sp, 89.75.Da}

\maketitle

\subsection{Motivations} 

 The way a granular bed responses to an applied shear stress reveals
 many of the peculiarities of this poorly comprehended "state" of
 matter~\cite{jaeger2000}. When a granular bed is sheared slowly
 enough by an elastic medium driven at constant velocity, nor the
 shear stress neither the shear rate can be directly controlled from
 outside \cite{annunziata16}. Rather, the system sets itself in a state at the edge
 between jamming and
 mobility~\cite{Dalton2001,pica10,dalton05,baldassarri06,petri08,geller2015,Zadeh18},
 exhibiting intermittent flow also called stick-slip.  This is an
 instance, among many others, of phenomena displaying intermittent and
 erratic activity, in the form of bursts, or {\it avalanches},
 characterized by wild fluctuations of physical quantities, and for
 this reason named {\it crackling noise}~\cite{sethna01}. Examples
 include earthquakes~\cite{main96}, fractures~\cite{petri94},
 structural phase transitions~\cite{cannelli93} and plastic
 deformation~\cite{dimiduk06}. These diverse phenomena share several
 common statistical features. In particular physical quantities
 display often long range correlations and self-similar
 distributions, i.e.  power laws, over a wide range of values. Such
 properties are usually ascribed to the vicinity of some critical
 transition~\cite{bak88,sethna01}, which in granular media could be
 the jamming transition~\cite{biroli2005}. Consistently, critical
 transitions bring about the existence of universality classes:
 systems that are microscopically very different, can display similar
 and universal statistical properties in their critical dynamics.
Within this spirit we have designed an experimental setup suitable to
observe such an irregular granular dynamics~\cite{dalton05},
characterized by critical fluctuations and reminiscent of that
displayed by the aforementioned wide class of physical systems.

In order to compare different systems exhibiting
critical dynamics, several quantities can be analyzed.   Recent literature witnesses a surge of interest for the average avalanche (or burst) shape (or profile). Introduced in the context of Barkhausen noise in ferromagnetic materials~\cite{Kuntz2000}, the average avalanche shape can provide a much sharper tool to test theory against experiments than the simple comparison of
critical exponents characterizing probability distributions.  As shown for simple stochastic processes, the geometrical and scaling properties of the average shape of a fluctuation depends on the temporal correlations of the dynamics~\cite{baldassarri:060601,colaiori:041105,colaiori08}. Such observation has allowed, for instance, to evidence a (negative) effective mass in magnetic domain walls~\cite{zapperi05}. 

Average avalanche shapes have been investigated for a variety of materials, well beyond magnetic systems~\cite{Papanikolaou2011}. Among the others: dislocation avalanches in plastically deformed intermetallic
compounds~\cite{chrzan1994} and in gold and niobium crystals~\cite{sparks2017}; stress drop avalanches at the yielding
transition in metallic glasses~\cite{antonaglia14} and, via numerical
simulations, in amorphous systems~\cite{ferrero16,Lagogianni2018};  bursts of load redistribution in heterogeneous materials under a constant external load~\cite{PhysRevLett.111.084302}.  Many biological studies have also measured average burst shape in cortical bursting activity~\cite{Roberts2014,Wikstro2015}, in transport processes in living cells~\cite{PhysRevLett.111.208102}, as well as in ants~\cite{Gallotti2018} and human~\cite{Chialvo2015} activity. 
Many other bursty dynamics have been investigated by means of this general tool, as stellar processes~\cite{PhysRevLett.117.261101} or  Earth's magnetospheric dynamics~\cite{Consolini2008}, and earthquakes~\cite{metha06}. The dependence  of the avalanche shape from the interaction range has been  studied  in elastic depinning models  ~\cite{laurson13}. 

In this paper we  acquire and analyze for the first time the average shapes of {\it slip velocity} and of  {\it friction force} in a sheared granular system, directly in the deep critical phase where it displays intermittent flow. Our findings also shed light on  apparently contradictory recent  observations \cite{bares17,denisov17}, and  supply new essential elements to improve the formulation of new and more effective dynamical models, with important impact on the understanding of related natural and technological issues.

\subsection{Introduction}

\begin{figure}[h]
\centering
\includegraphics[width=\linewidth]{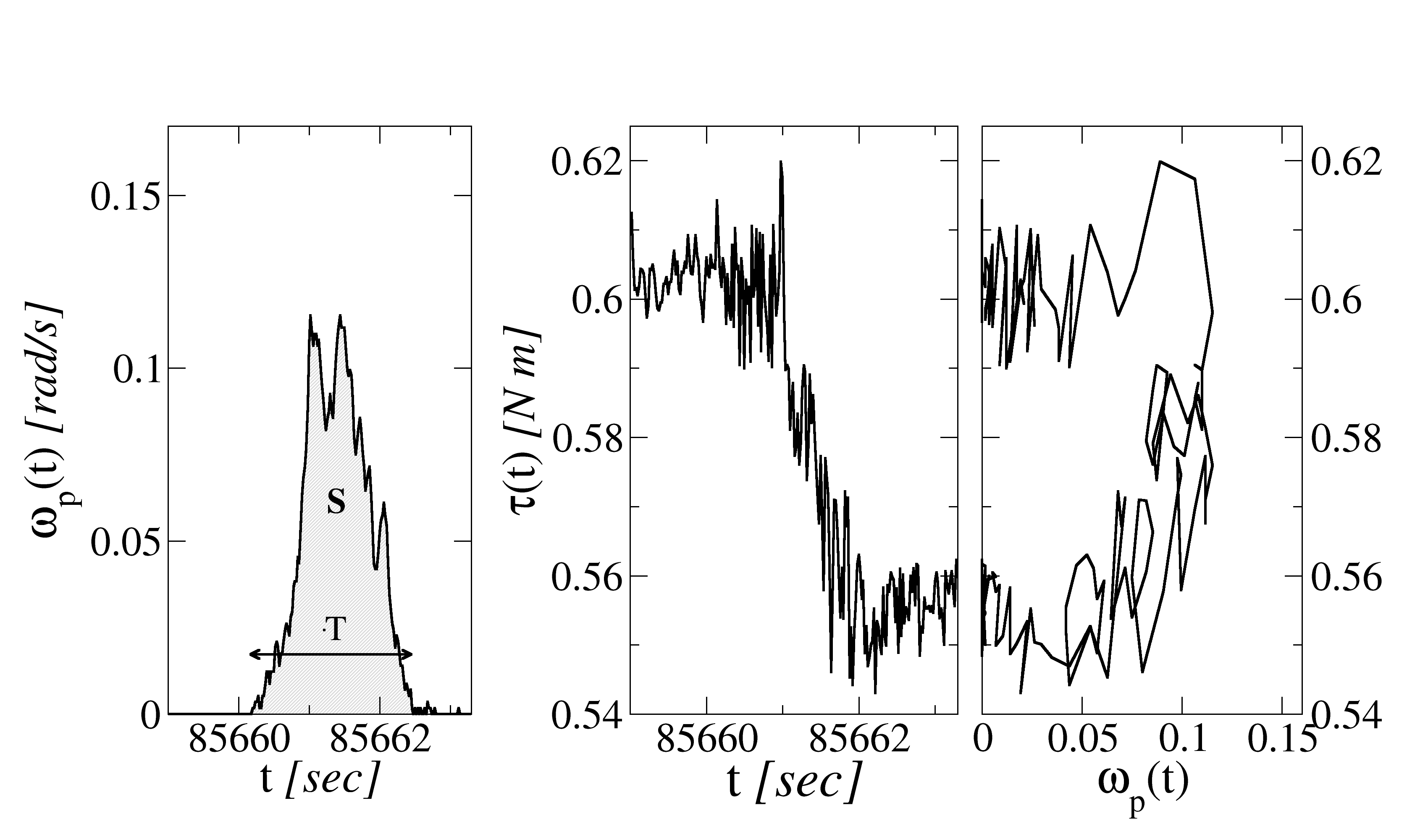}
\caption{Sample of raw data for a slip event. Left: instantaneous velocity of the slider versus time. Upper axis of the graph reports the total time elapsed from the beginning of the experiment, while bottom axis indicate the internal avalanche time, starting from $0$ when the slip begins, and ending at slip duration $T$. The area below the curve is the total slip size $S$. Center: Friction torque experienced by the slider in the same time window. Right: Friction torque vs instantaneous slider velocity in the same time window.}
\label{fig1}  
\end{figure}

We study the stick-slip dynamics at the level of the single slip event, as illustrated in Fig.~\ref{fig1}.
The left panel reproduces the angular velocity, during a slip, of a slider that rotates while in contact with the granular bed. The middle panel shows the corresponding frictional torque experienced by the slider. 
The motion can be described as a function of the {\it internal avalanche time} $t$, which starts at the beginning of the slip and ends when the system sticks. Each slip event has its own duration $T$ and its size $S$ (the grey area in Fig.~\ref{fig1}, left panel). The average velocity shape is performed considering many slips with the same total duration $T$ as function of internal avalanche time $t$.  A similar averaging procedure is followed to obtain the average friction shape: i.e. the average friction torque exerted  by the granular medium at the internal time $t$ during a slip event of total duration $T$.  The right panel of Fig.~\ref{fig1} shows the intricate, complex relation between friction and velocity during the intermittent, stick-slip dynamics. 

In our study we observe the existence of a cross-over from small to large slips. We identify it as a breakdown of the critical scaling and show that such transition is in turn related to a change in the frictional properties of the system. Specifically, we find that the average velocity of the cross-over avalanches corresponds to a characteristic value marking a dynamical transition from weakening to thickening frictional behavior of the system. 
Average  shape for avalanches of stress drop~\cite{denisov17} and energy drop rate~\cite{bares17} have  been recently investigated in slow but continuous flow, where velocity never drops to zero and the stress is the relevant fluctuating quantity.    While in ~\cite{denisov17} average avalanches  have been found to display symmetrical and self-similar shapes, in \cite{bares17} these properties have been observed only in  avalanches sufficiently small.  Our investigations, conducted in the critical state,   contribute to clarify the origin of these contradictory behavior observed in a different situation.

\subsection{Experimental set up} 

The experimental set up (see Fig.~\ref{fig-expsetup}) is similar to that
employed in~\cite{dalton05,baldassarri06,petri08,leoni10,pica12}
and described in more detail in the Appendix.
The apparatus consists of an assembly of glass spheres
laying in an annular channel and sheared by a horizontally rotating top
plate driven by a motor.  The instantaneous angular position of the
plate and of the motor, respectively  $\theta_p$ and $\theta_0$ are  acquired 
by means of two optical encoders.

The plate is  coupled to the motor  through a soft torsion spring of  elastic constant $k$, so the  instantaneous frictional torque, $\tau$, exerted by the granular medium  can be derived from the equation of  motion for the plate:
 \begin{equation}
 \label{motion}
\tau = -k (\theta_0-\theta_p) -  I \ddot{\theta}_p,
\end{equation}
where $I$ is the inertia of the plate-axis system.  The motor angular speed $\omega_0$  is kept constant, so that $\theta_0(t)=\omega_0 t$,  but the interaction  between the top plate and the granular medium  is crucial in determining the instantaneous plate velocity, leading to different possible regimes.  When both the driving speed and spring constant are low enough the critical dynamics, in which the plate  performs highly irregular and intermittent motion, is approached.

\subsection{Scaling analysis}

\begin{figure}[h]
 \centering
  \includegraphics[width=\linewidth]{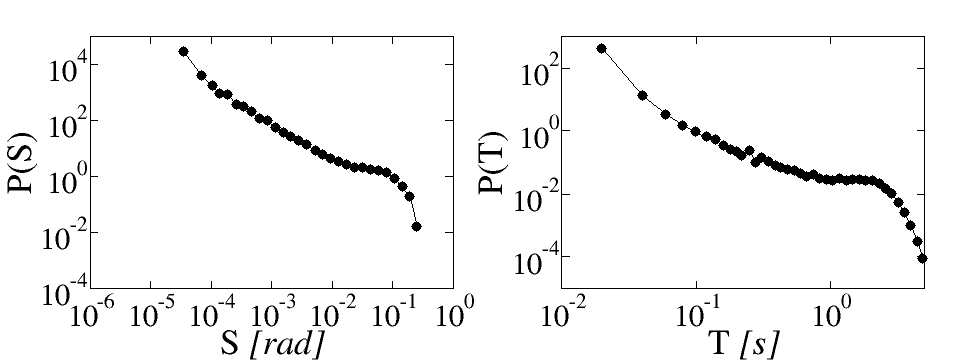}
 \caption{Left: Probability distribution of slip extensions ($S$). Right: Probability distribution of slip durations.}
 \label{fig2}  
 \end{figure}

We have performed long experimental runs in the critical, stick-slip, regime measuring the angular coordinate of the plate $\theta_p(t)$, from which we have derived the plate angular velocity $\omega_p(t)$. We have collected statistics for a large number of avalanches: the distribution of corresponding durations $T$ and sizes $S=\int_0^T  \omega_p(t) dt$ are shown in Fig.~\ref{fig2}. Both distributions exhibit a slow decay, roughly close to a power law, terminating by a bulging cutoff for large sizes. Similar broad distributions are shared by other quantities, e.g. the plate velocity, $\omega_p$~\cite{baldassarri06} (not shown).
 As recalled in the introduction, power law decay in distributions are generally considered the hallmark for criticality. If this is the correct scenario, one should observe self-similar scaling relations in average quantities too. In particular, we consider the average shape of velocity during an avalanche of a fixed duration, defined as:
\[
\langle \omega_p(t)\rangle_T = \frac 1{N_T} \sum_i \omega^{(i)}_p(t),
\] 
where $\omega^{(i)}_p$ is the plate velocity during the $i_{th}$ observed avalanche of duration $T$, whose total number is $N_T$, and $t$ is the internal time within the slip: $0<t<T$. 
Although the average velocity shape $<\omega_p(t)>_T$ depends on both $t$ and $T$, criticality should imply that an invariant function  $\Omega$ exists, such that it  can be expressed as:
\begin{equation}
<\omega_p(t)>_T = g(T) \Omega(t/T).
\label{scaling}
\end{equation}
The function $g(T)$ determines how the average event  size $<S>$ scales with respect to
the slip duration $T$. In fact, integrating the above equation  with respect to $t$ one gets:
\begin{equation}\label{eq-2}
<S>_T \, = T g(T)
\end{equation}
(where without loss of generality we have assumed $\int_0^1 \Omega(x) dx = 1$). The function $\Omega$ represents the average invariant
pulse shape, which is expected not to depend on the slip duration  and  can be computed via the above equations as
\begin{equation}
\Omega(t/T) = T\,\frac{<\omega_p(t)>_T}{<S>_T}.
\label{faverage}
\end{equation} 

The previous scaling scenario is produced by several theoretical models for critical dynamics. One paradigmatic model for crackling noise is the so called ABBM model~\cite{Alessandro1990}, proposed to describe the intermittent statistics of electric noise recorded during hysteresis loops in ferromagnetic materials (Barkhausen noise). It is simple enough to allow exact analytical results~\cite{Alessandro1990,Feller1951,colaiori08,Papanikolaou2011,Dobrinevski2012}, and it predicts power law distributions for avalanche sizes and durations, as well as parabolic average avalanche shape, in the scaling regime. In the conclusive section we will discuss the connections between this model and what we observe in our study.

\begin{figure}
\includegraphics[width=\linewidth]{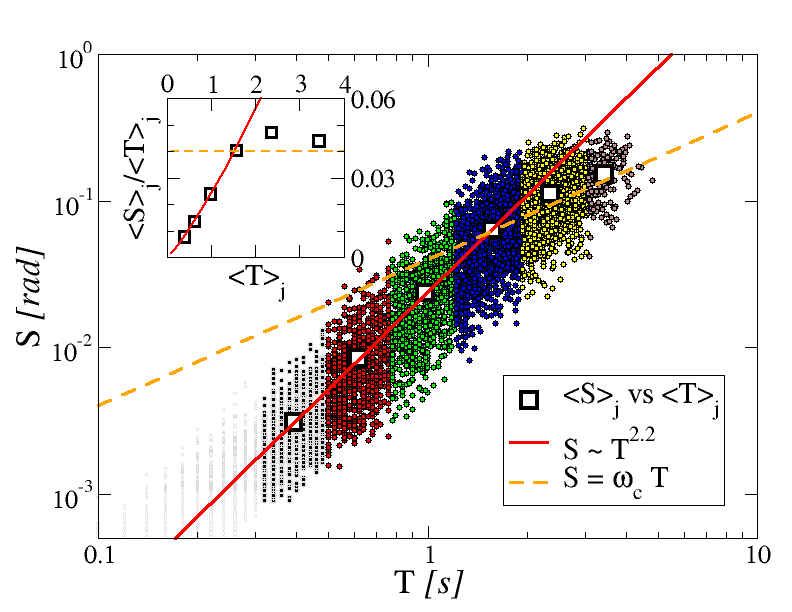}
\caption{\label{fig3} Scatter plot of size $S$ vs duration $T$ of each single slip. Symbols (colors online) correspond to the different duration classes employed for the average shape analysis. Inset: Average slip velocity $\langle S \rangle_j/\langle T \rangle_j$ for each class as a function of average duration $\langle T \rangle_j$ of the class. Lines (both in main plots and in inset), are guide to the eyes for: power law behaviour $S\approx T^{2.2}$, and linear behaviour $S = \omega_c T$ (where $\omega_c=0.04$, see text and Fig.~\ref{fig5} for definition).}
\end{figure}

\subsection{Average shape of slip velocity}

To investigate the properties of the average pulse shape, and to test its invariance and the scaling hypothesis,   we have divided all the avalanches observed in the experiments into classes according to their  duration (see  Appendix). Figure~\ref{fig3} (main panel) shows the slip size as function of its duration for all the slips considered in the statistics, and the different colors correspond  to the different classes of duration. For each class, $j$, we have computed the average  slip size $<S>_j$ and  duration $<T>_j$, and 
the average velocity $<\omega_p(t)>_j$ measured as function of the internal time $t$.  According to Eq.~(\ref{faverage}), in order to obtain $\Omega(t/T)$, this average velocity has then been normalized to the ratio $<S>_j/<T>_j$.

\begin{figure}
\begin{center}
\includegraphics[width=\linewidth]{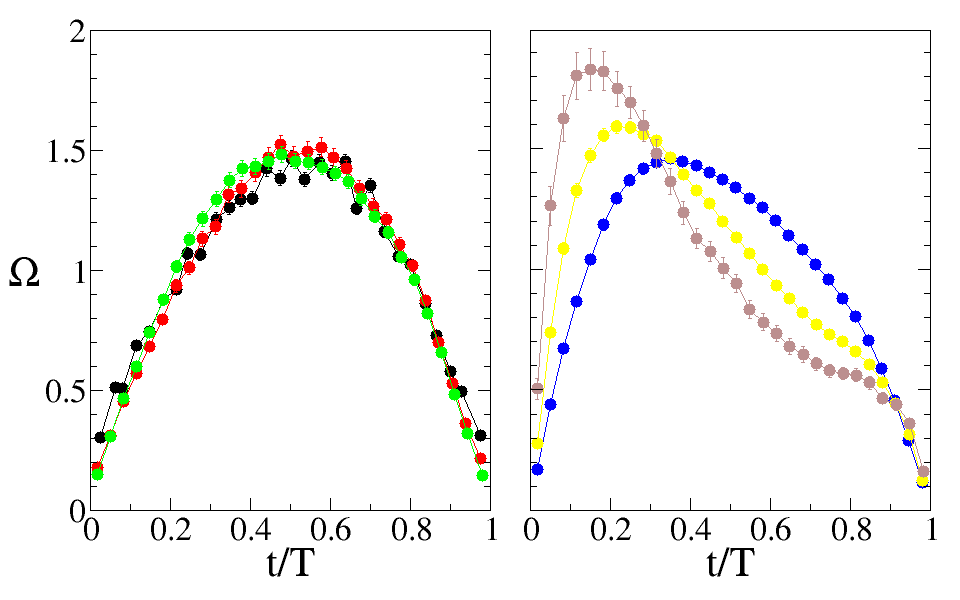} 
\end{center}
\caption{\label{fig4} Average velocity  profile of slips from experiments
  rescaled  by their duration $T$, and size $S$ according to Eqs.~(\ref{scaling}) and~(\ref{faverage}). Different curves correspond to different slip duration ranges. Colors refers to the duration class, as shown in Fig.~\ref{fig3}. Left panel shows classes of ``short'' avalanches, right panel classes of ``large'' avalanches (see also Appendix). }
\end{figure}

The resulting average shapes for each class of duration are shown in Fig.~\ref{fig4} (light, grey points in ~\ref{fig3}, corresponding to very short slips at the limit of the system resolution, have been discarded).  All classes exhibit comparable values of the rescaled maximum velocity implying that longer avalanches are also faster.  However, rescaled average shapes unveil that there are two kinds of avalanches. Some of them,  say {\it short}, have the  shape  described by a unique function $\Omega(t/T)$, visible in Fig.~\ref{fig4} (left panel). That is, their  size and duration are related by the well defined scaling law Eq.~\ref{scaling}.  On the contrary, the average velocity shapes of {\it long} avalanches (right panel) change with the duration and cannot be reduced to a universal form by a  homogeneous rescaling of the variables. Moreover, they do not display the almost symmetric shape characterizing small avalanches.

As anticipated, Bar\'es and coworkers~\cite{bares17} have recently measured the average shape of stress drop rate avalanches  in a bidimensional granular system driven at constant shear rate. Similarly to the present findings, they observe  that larger slips develop left asymmetries. They have  hypothesized a possible role of the static friction between particles and supporting glass, and of nonlinear elasticity, given by the relatively soft nature of the grain material employed in their experiments. We can however exclude these factors in our experiments, where the interface grain-wall is small with respect to the bulk and the beads are made of glass.
The leftwards asymmetry observed in experiments represents a very interesting phenomenon, which in general is expected from non trivial dynamical effects, and cannot be due to the simple inertia of the moving plate (which should produce opposite asymmetry~\cite{baldassarri:060601,colaiori:041105}). 

In some magnetic materials, a leftwards asymmetry has been observed and related to memory effects acting as an effective negative mass of domain walls~\cite{zapperi05}. In our experiments we cannot exclude the existence of such an ''effective'' inertia of the system, due to  some memory introduced by the underlying granular. For instance, in some  experiments~\cite{nasuno98} researchers noted an increased inertia of the slider moving on a granular bed, due to the grains dragged by the slider itself and  a similar augmented inertia has been observed also in our previous experiments~\cite{baldassarri06}. Since the quantity of grains dragged by the disk during its motion could change during the irregular motion of the system  one should consider the inertia as a dynamical quantity, rather than a constant, and this could in principle be one origin of the asymmetries. A left asymmetry has been also observed in earthquakes~\cite{Houston1998,metha06}.

\subsection{Breaking of scaling}
 
More insight into the mechanisms leading to the scaling breakdown can be gained by looking again at the  plot relating  $S$ and $T$ shown in Fig.~\ref{fig3}. The first information coming  from this plot is that there exists a definite statistical scaling between slip size and duration, as shown by the scattering of data. The white squared symbols in the main plot represent the average slip size and duration of each class (statistical errors  are negligible on these averages). It is seen that, at least for the four lower classes, they follow an algebraic relation: $<S>_j  \simeq <T>_j^{\alpha}$ (red continuous line).  The value of the exponent turns out to be  $\alpha \simeq 2.2$. This is close to, but clearly different from, the value of $\alpha = 2$ expected from extant models (for instance the ABBM model mentioned above~\cite{Alessandro1990}).

The other information supplied by the scatter plot of Fig.~\ref{fig3} is that this behavior changes at large slips, where a linear dependence, $<S> \propto <T>$ looks more appropriate (yellow dashed line). Interestingly, the crossover between the two behaviors takes place around the fourth class, exactly where the scaling of the average pulse shape, shown in Fig.~\ref{fig4}, breaks down. 
The inset of Fig.~\ref{fig3} puts into better evidence this cross-over. There, we have plotted the quantity $<S>_j/<T>_j$ as function of $<T>_j$.  We observe a weakly superlinear relation for small slips, followed by
a plateau at large slips. Note that the ratio between $S$ and $T$ is nothing but the plate average velocity  during the slip.  This observation allows to identify a critical velocity, as the ratio between the  average slip size, $<S_c> \approx 0.063 $ rad,  and the average duration, $<T_c> \approx 1.57$ s,  of the fourth class: $\omega_c = <S_c>/<T_c> \approx 0.04$ rad/s. We speculate that during large slips ($T>T_c$), when the plate reaches high velocities $\omega_p>\omega_c$, it could experience some sudden increase of friction.
In the next section it will be seen that this increase indeed appears, as a dynamical effect. 

\subsection{Stochastic friction}

The forces ruling the slip dynamics are the spring torque and the granular friction. While the first one
just depends linearly on the instantaneous angle, the second displays a complex behavior (see Fig.~\ref{fig1}, central and right panel) from which interesting features emerge.

The classical Mohr--Coulomb criterion predicts constant friction at low shear, and increasing values when the system enters the Bagnold's regime~\cite{bagnold54,bagnold66}, a behavior well observed experimentally at constant shear (see e.g. \cite{savage84}).  
  \begin{figure}
 \includegraphics[width=\linewidth]{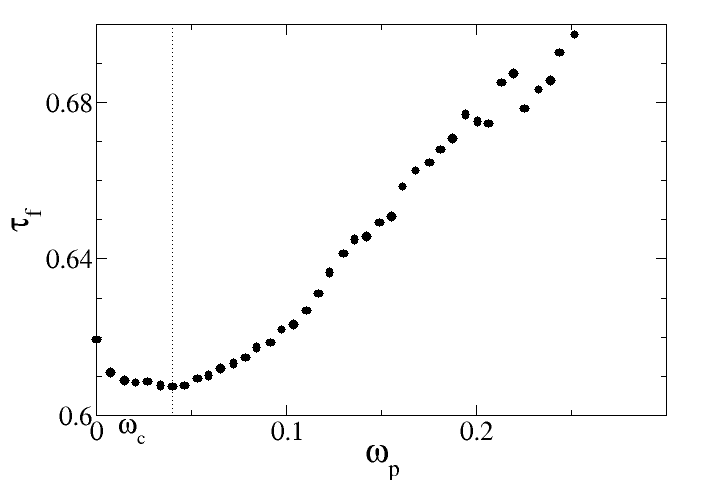}
 \caption{\label{fig5}
   Conditioned average friction torque (see Eq.~\ref{avefric}) as a function of the instantaneous plate velocity in experiments
 }
 \end{figure}
 However, it is doubtful whether this behavior could be relevant to the stick-slip dynamics observed in the critical regime.   More generally, friction in granular systems is usually measured under controlled shear strain or stress, but its properties can be dramatically different when observed in the self-organized state, as exemplified in Fig.~\ref{fig1}, right panel.  Some statistical features of friction in this state have been investigated in~\cite{dalton05,baldassarri06,petri08}, but despite this quantity plays a crucial role for the system dynamics,  it has never been systematically measured to date during stick slip. 

In the critical regime  friction is a random quantity. It depends on the details of the network of contacts 
 between particles in the granular bed. Fluctuations in the frictional response of the granular medium result from the stress propagation on the evolving network of grain contacts, and are at the very origin of the motion stochasticity.
This fact has a number of consequences and some subtleties. A random friction force, as a stochastic quantity, can be described by statistical estimators like  averages, moments, correlators, etc. Nevertheless, several averages can be defined, which depend on the driving protocol and can be very different from each other. 
More specifically, one can consider the time average of the friction over the full dynamics, but this is not always really meaningful, especially in the critical regime. As shown in~\cite{dalton05} the statistical distribution of friction in this regime is characterized  by fat tails, as opposite with  the continuous sliding where it is normal. Another possible average,~\cite{baldassarri06,leoni10} is the average friction {\em conditioned} to the (instantaneous) plate velocity:

 \begin{equation}
 \label{avefric}
 \tau_f(\omega) =  \lim_{T\to \infty} \frac{\int_0^T \tau(t) \delta(\omega-\omega_p(t))dt}{\int_0^T \delta(\omega(t)-\omega_p)dt}.
 \end{equation}
 
In  Fig.~\ref{fig5}   we plot such conditioned  average friction during the stick slip critical regime.
As noted in~\cite{baldassarri06}, an interesting Stribeck-shaped (that is, a shear weakening followed by a thickening)  friction curve appears, featuring weakening for small velocities and recovering the Bagnold behavior at high velocities. However, this velocity weakening arises as a dynamical effect. In fact, a different driving protocol can give different results: For instance,  at constant shear~\cite{savage84} the average friction is constant at low and intermediate speeds.

The analysis of Fig.~\ref{fig5} allows to identify a velocity corresponding to the position of the minimum of the average friction $\tau_f$. Our experiments clearly indicate that the position of this minimum does not depends on the drive velocity (Fig.~\ref{Fig6SeveralDrives} in Appendix) and it is always attained near the velocity $\omega\approx 0.04$ rad/s. This value is very close to the value $\omega_c$ marking the crossover in the scaling of $S$ vs $T$ (see Fig.~\ref{fig3}), which in turn is related to the breakdown of scaling of the average avalanche shape shown in Fig.~\ref{fig4}.
This corroborates the previous interpretation of the crossover phenomena and the breaking of the critical scaling of the dynamics as due to the weakening followed by the increase of friction experienced by the plate during larger, faster avalanches Fig.~\ref{fig5}.

\begin{figure}
\begin{center}
\includegraphics[width=\linewidth]{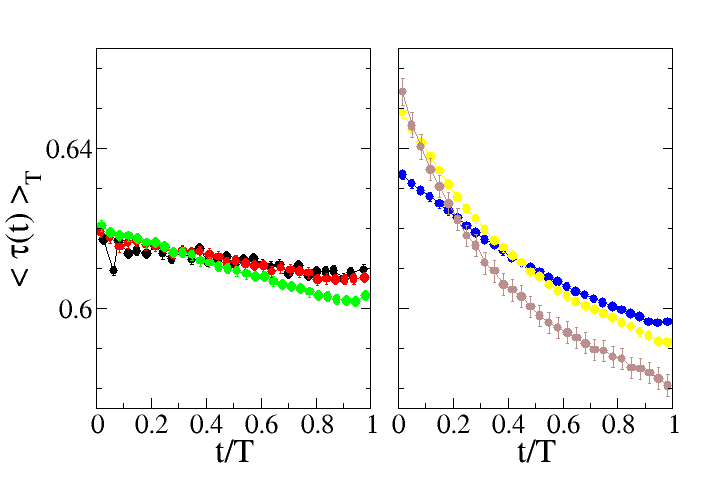} 
\end{center}
\caption{\label{fig6} Average friction torque along slips of different duration as function of  rescaled time in experiments.
  Colors refers to the duration class, as shown in Fig.~\ref{fig3}.  Left panel shows classes of ``short'' avalanches, right panel classes of ``large'' avalanches (see also Appendix).}
\end{figure}

In order to better investigate  whether and how friction dynamical behavior can influence the average velocity shape we have also analyzed the average shape of friction along the slip. 
In an analogous fashion to what done for computing the velocity shapes, one can define $\langle \tau(t)\rangle_T$ as the average frictional torque for slips of the same duration $T$. In practice, we have computed the average value of the friction torque over slips of similar duration $T$, according to the same classes of duration adopted for velocities (see  Fig.~\ref{fig3} and Appendix). 
The results, presented in Fig.~\ref{fig6}, show  that the breaking of scaling of the velocity shapes corresponds to a change in the frictional properties of $\langle \tau(t)\rangle_T$. 
For small avalanches, i.e. those corresponding to the cases in which  average velocity shape  obeys scaling (curves plotted in the left graph of Fig.~\ref{fig6}),
the average friction maintains  an almost constant value along the whole slip, whose value is independent fron the slip duration.
On the contrary,  the curves corresponding to longer slips (shown in the  right t plot of Fig.~\ref{fig6}) display different shapes that,  as in the case of  velocity (Fig.~\ref{fig4}), strongly depend on $T$, and cannot be collapsed. Note also that higher frictions are experienced during longer slips.

Let us stress here the difference between the two average procedures considered in this work. The average $\langle \tau \rangle_T$ shown in Fig.~\ref{fig6}, are performed over slips of similar duration, at the same internal avalanche time $t$. Instead, the (conditional) average $\tau_f$, defined in Eq.~\ref{avefric} and  shown in Fig.~\ref{fig5}, mixes events of any duration, and it depends on the instantaneous plate velocity $\omega_p$. The two quantities give different aspects of the same (stochastic) physical phenomenon.
Nevertheless, the combination of the two analysis suggests that the quite complex friction weakening behavior of $\tau_f$ is mainly due to large slips, which show a non constant average friction $\langle \tau \rangle_T$ in time (see Fig.~\ref{fig6}, right panel), in contrast with small avalanches, where the average keeps mainly constant.

\begin{figure}
\begin{center}
\includegraphics[width=\linewidth]{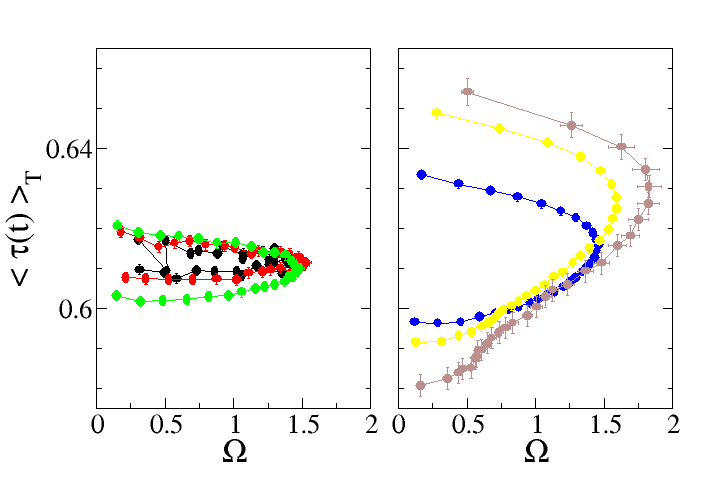} 
\end{center}
\caption{\label{fig7} Average friction torque along  slips  of different duration as function of the normalized average slip velocity in experiments
  Colors refers to the duration class, as shown in Fig.~\ref{fig3}.  Left panel shows classes of ``short'' avalanches, right panel classes of ``large'' avalanches (see also Appendix).}
\end{figure}

\begin{figure}[h]
  \begin{minipage}{0.4\textwidth}
   \includegraphics[width=\textwidth]{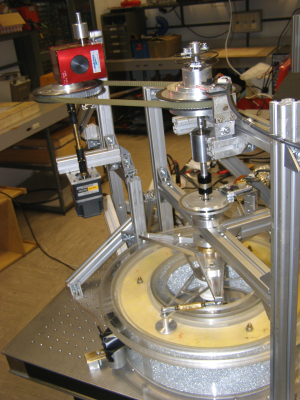}
   \end{minipage}
  \hfill
  \begin{minipage}{0.5\textwidth}
    \includegraphics[width=\textwidth]{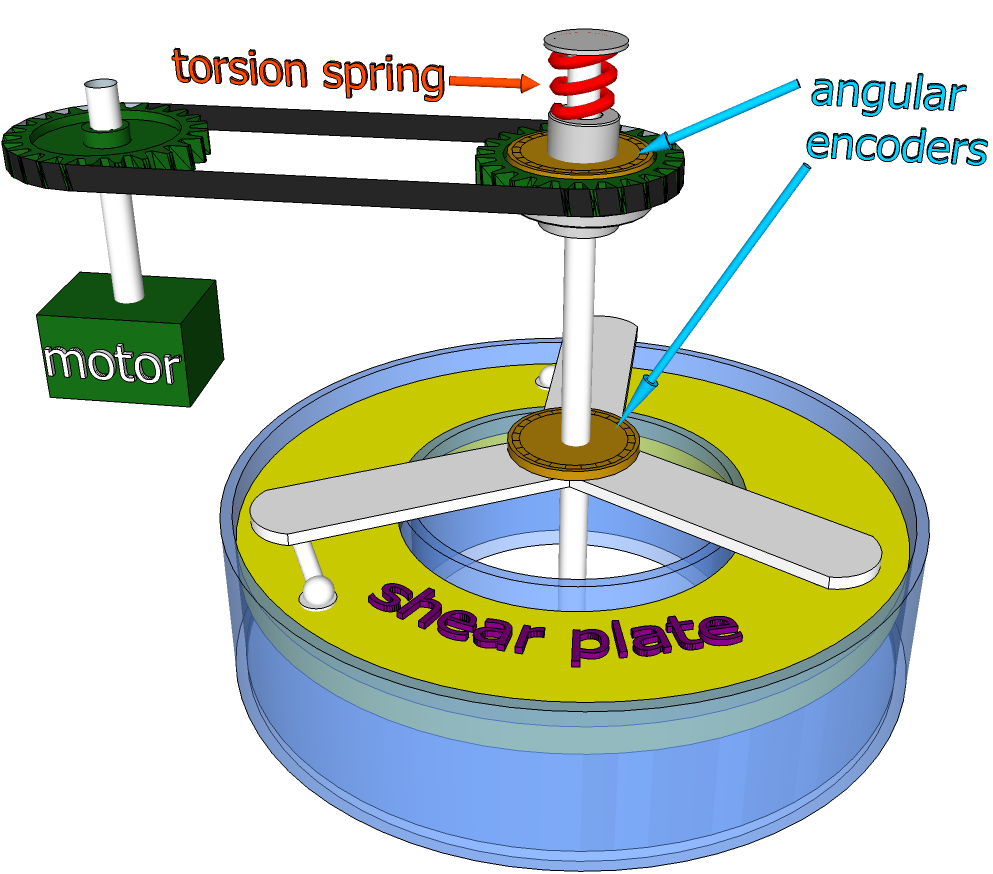}
   \end{minipage}
  \caption{Photo (up) and schema (down) of the experimental set up.}
 \label{fig-expsetup}
  \end{figure}

By combining the analysis of friction and velocity shapes, one can consider 
the curves resulting from plotting the average friction as a function of the average velocity,  in slips of similar duration, as shown in Fig.~\ref{fig7}. 
They  show that while, as anticipated, friction has a low velocity dependence in small slips (left panel),  in large ones it splits into a two-valued function (right panel), 
displaying an hysteresis, with  well different dependencies on the (average) plate velocity  in the accelerating and decelerating phases of the slips. 
This evolution  is  very similar to that observed in periodic stick-slip~\cite{nasuno97,nasuno98},   where  all slips have identical extension, duration, and velocity profile.

\subsection{Considerations and conclusions}

Our experiments show a good scaling  of the average velocity  shape for small avalanches, with an almost symmetric average shape.
For larger avalanches however, scaling~(Eq.~\ref{scaling}) is broken:
for large slips the shape takes a clear leftwards shape in agreement with what observed in seismic data~\cite{Houston1998,bares17} (and recently in~\cite{bares14}).

Our analyses show that the breakdown of velocity scaling  goes along with changes in the friction behavior,  pointing  out a strict relation  between the two phenomena. On the opposite, spring-block models with only Coulomb friction  generate symmetric slips~\cite{aharonov04,bizzarri16}.  Effective friction laws accounting  for elapsed time and/or  space have been incorporated  in solid-on-solid interface models, through  the  dependence on so called state variables~\cite{ruina83,dieterich94,baumberger06}. 
 These {\it rate-and-state} laws are often  adopted for studying and modeling co-seismic fault shearing, together with their other simpler forms \cite{scholz98,bizzarri2003,kawamura2012,bizzarri16}. They are essentially phenomenological  and can describe both velocity weakening and hardening, depending on  the adopted parameters (which are not derivable from microscopic principles). These laws 
  have shown to  work to some extent also for interstitial granular matter \cite{marone98}, but with some inconsistencies \cite{mair99}. Moreover they have been drawn from experiments where velocity is forced to change in sudden steps and   they don't seem to have been never investigated  in the critical stick-slip. Attempts to do this  with smoothly varying velocity have been done in	\cite{leoni10}. In  a  very recent work  \cite{degiuli17} both friction  weakening and hysteresis
  have been numerically investigated during granular shear cycles, showing that these are due to  contact instabilities induced by the acoustic waves generated during granular fluidization.  It is thus clear that granular flow cannot be effectually modeled without the inclusion of more refined and realistic friction laws.

An effective modeling approach to the critical granular dynamics cannot as well exclude a stochastic description of friction, which generates the slip unpredictability and their range of variability,  with the following change in the slip shapes. To our knowledge, the only few attempts in this direction~\cite{baldassarri06,leoni10,dahmen11} are  inspired to the aforementioned {\it ABBM model}~\cite{Alessandro1990}, which represents the mean field approximation for the motion of a driven elastic interface in a random environment~\cite{Zapperi1998,LeDoussal2012}.
From the dynamical point of view, it describes a spring--slider model subjected to a friction where both viscous and a random pinning components are present, in the overdamped (i.e. negligible inertia) approximation. At small, but finite driving rate, the ABBM model predicts an intermittent, critical dynamics for the block motion. Similarly to our observations, avalanche statistics show a scaling regime for short slips, whose average velocity has parabolic shapes.
However, an exponent $\alpha =2$ relates  $\langle S \rangle$ to   $\langle T \rangle$, which is different from what observed in our experiment. Moreover, for longer slips, ABBM predicts flatter symmetical shapes, witnessing a cut-off in the velocity correlation. No inertial effects are present, due to  the overdamped approximation. 

A variant of the ABBM model for critical granular dynamics has been introduced in~\cite{baldassarri06}, where, based on empirical observations, a simple Stribeck-like rate dependence, showing a minimum,  of the average granular friction was adopted.  Moreover, more physically,   a cut off in the spatial correlation of the random  force was considered  and, at odds with the original model, inertia was taken in account. Later on, attempts to introduce in the model a state dependent weakening friction have been  done~\cite{leoni10}), and further investigations are in progress. 

We think that the insights provided by the present study can explain the contradictory recent observations in \cite{bares17,denisov17} and can be useful to advance such  efforts to improve models. In particular, they show that inertia can play an important role in both weakening-hysteresis~\cite{degiuli17} and in the determining scaling exponent $\alpha$ (an inertial ABBM model has  been studied in~\cite{LeDoussal2012inertial}). 
Even at the microscopic level, grain inertia can influence the avalanche statistics. For instance, in sandpile models, largely studied in the context of SOC (Self Organized Criticality), the tendency of real sand grains to keep moving once they start  facilitate  the emergence of huge avalanches. Recent theoretical developments propose, in the presence of such facilitation effects, a scenario called Self-Organised Bistability~\cite{DiSanto2016}, where again a breaking of scaling is associated to the appearance of large avalanches ("kings").

We conclude that weakening is a genuine property exhibited by granular dynamics at variable shear rate, and that randomness and memory are a general features of friction that cannot be overlooked in the formulation of effectual models. Such models can have impact on the understanding of many phenomena occurring in the large realm of granular systems, 
and in particular of self organized natural phenomena like landslides, and earthquake,  where it is not yet clear the way different mechanisms can contribute to the shear weakening observed in coseismic fault shearing~\cite{ditoro11}. 
Further investigation on theoretical models incorporating  more realistic, specifically memory dependent friction laws, and new experiments will allow to better understand the mechanism for which criticality breaks down.

\section*{Acnokwlegments}{This work has been supported by the grant FIRB RBFR081IUK\_003
from the Italian Ministry for Education and Research 
}

\section[A]{Appendix}

\subsection{The experimental set up}

The experimental apparatus utilized for this research consists of a
circular PPMI channel of outer and inner radii $R$ = 19.2 cm and $r$ =
12.5 cm respectively.  The channel is 12 cm height and is almost
filled with a  bidisperse mixture 50\%-50\% of glass beads,  with radii $r_1$=1.5
mm $\pm$ 10\% and $r_2$=2 mm $\pm$ 10\%. 

A top plate, fitting the channel, can be rotated and has a few layer of grains glued
to its lower face in order to better drag the underlying granular medium. The plate has mass $M$ = 1200 g and moment of inertia $I$ = 0.026 kg
m$^2$, and it is free to move vertically, implying that in our experiments
the medium can change volume under a nominal pressure of $p= M g /
[\pi (R^{2} - r^{2} )] \approx $176 Pa. The plate is connected to an end of torsion spring, of elastic constant $\kappa =$ 0.36 Nm/rad, while the  other end of the spring is rotated by a motor at constant angular velocity $\omega_d$. 
  The angular positions of
motor and plate are supplied by two optical encoders  positioned on
either side of the torsion spring, each one having a spatial
resolution of $3 \cdot 10^{-5}$ rad and  being sampled at 50 Hz. These measures provides the plate instantaneous position and velocity, 
$\theta_p$ and $\omega_p$,  as well as the friction torque, which is proportional to the angular difference between motor and plate(see Eq. (\ref{motion})).

\subsection{Experimental analysis}

In principle each single slip event, or {\it avalanche}, begins when
$\omega_p$ starts to differ from zero and ends when $\omega_p$  goes
back to zero. However, in practice it is necessary to choose a
threshold value $\omega_{th}$ to cross, in order to get rid of the
instrumental noise. This choice is to some extent arbitrary, however
all the results have been observed to be independent from
the chosen threshold, as long as it is small and different from 0. For our analysis  we have set $\omega_{th}=0.00175$ rad/s, and considered the seven time series reported in table~\ref{tablediffdrives}.

 \begin{table}
 \begin{center}
   \small
   \begin{tabular}{c|cccccc}
 series 
  & duration &\# of points  & driving $\omega_d$  & \# of slips  \\
  &  
  (minutes)  & & (rad/s)  &  used in analysis  \\
 \hline
 $(EA)$ 
 & 3900  & 5849962 & 
 0.0015 &  6014 \\
 $(EB)$ 
 & 673  & 2020079 & 
 0.0022 & 1625 \\
 $(EC)$ 
 & 1200 & 3600060 & 
 0.0044 & 5826 \\
 $(ED)$ 
 & 4080  & 12240020  & 
 0.0055 & 2451 \\
 $(EE)$ 
 & 360  & 1079977 & 
 0.011 & 3725 \\
 $(EF)$  
 & 240  & 720007 & 
  0.021 &  3973\\
 $(EG)$ 
 & 210  & 630014 & 
 0.033 & 4300 \\
  \end{tabular}\\
 \caption{\label{tablediffdrives} Features of the analyzed series of experiments with different drives}
 \end{center}
 \end{table}

The avalanches of each series have been grouped into classes on the base of their duration, according to the first column of table~\ref{tableclassesEA}.
For each class $j$ the average avalanche duration $<T>_j$ and size $<S>_j$ have been evaluated, and instantaneous average velocity has been computed at a set $\{t_i\}$ of discrete times,  $0 \le t_i \le~<T>_j$ (see main text).  

Avalanches at the extremes of distributions have been dropped out.
For example duration and size for avalanches from the series (EA) are plotted in Fig. 2 of the main text, with  different colors for each interval. Avalanches in gray, shorter than 0.31 s, are too small to perform meaningful analysis (less than 15 points at $50$Hz of sampling rate).  The total number of
avalanches employed in this statistics has then been 6014, distributed according to the second column of table~\ref{tableclassesEA}.  

\begin{figure}[]
\begin{center}
\includegraphics[width=\linewidth]{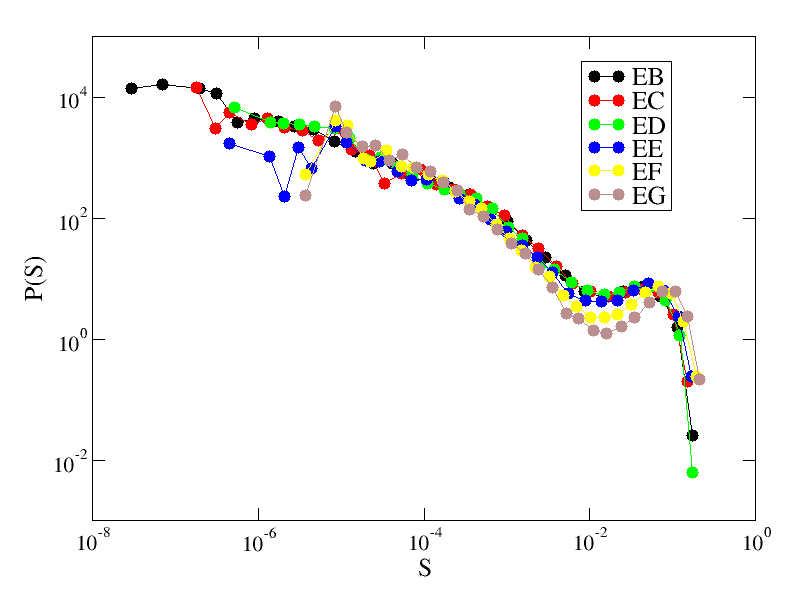}
\caption{Avalanche size distributions for different drive velocities (see Fig.~\ref{fig3} in the main text). }
\label{Fig2SeveralDrives}
\end{center}
\end{figure}

\begin{figure}[]
\begin{center}
\includegraphics[width=\linewidth]{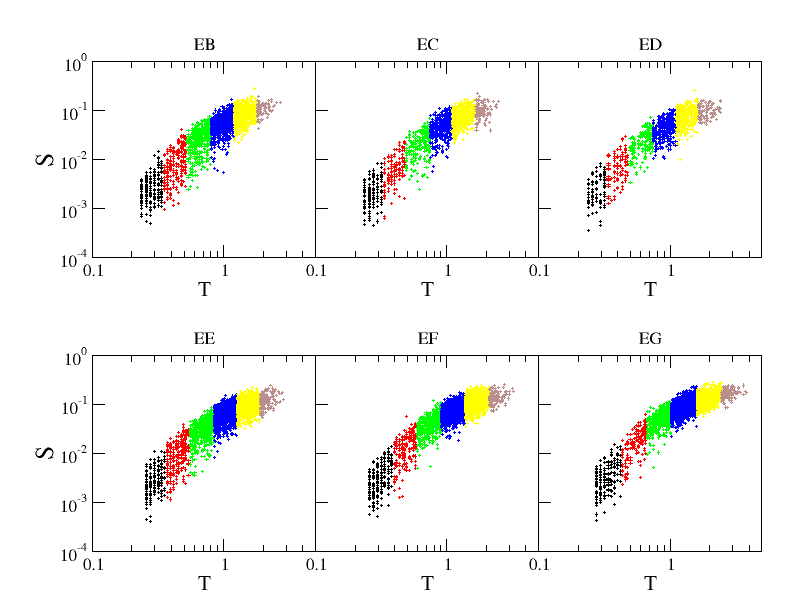}
\caption{Avalanche sizes vs durations, and class definitions, for different drive velocities (see Fig.~\ref{fig2} in the main text) }
\label{Fig3SeveralDrives}
\end{center}
\end{figure}

\begin{figure}[]
\begin{center}
\includegraphics[width=\linewidth]{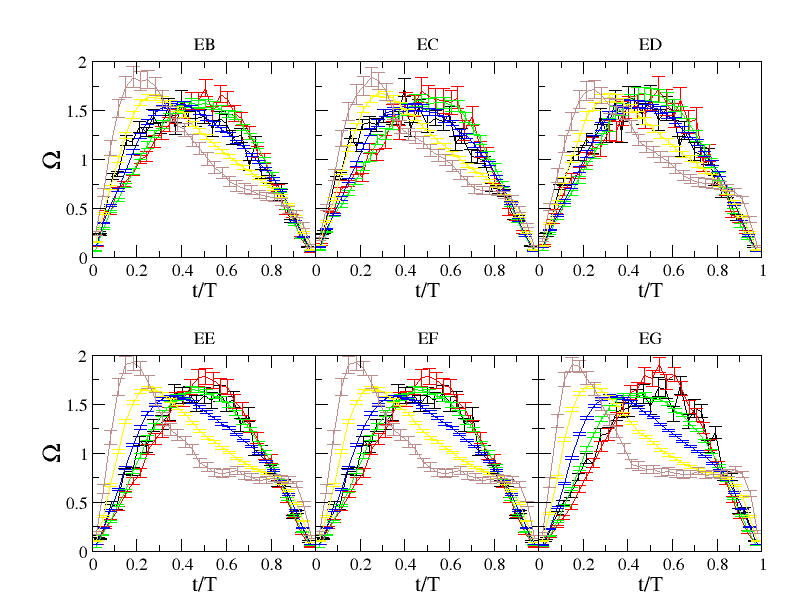}
\caption{Average velocity shapes for different drive velocities (see Fig.~\ref{fig4} in the main text) }
\label{Fig4SeveralDrives}
\end{center}
\end{figure}

\begin{figure}[]
\begin{center}
\includegraphics[width=\linewidth]{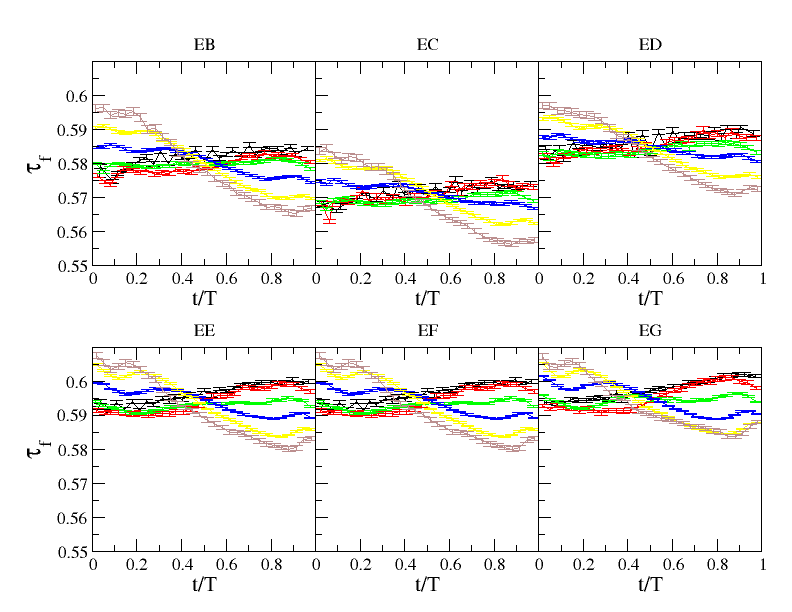}
\caption{Average friction shapes for different drive velocities (see Fig.~\ref{fig5} in the main text) }
\label{Fig5SeveralDrives}
\end{center}
\end{figure}

\begin{figure}[]
\begin{center}
\includegraphics[width=\linewidth]{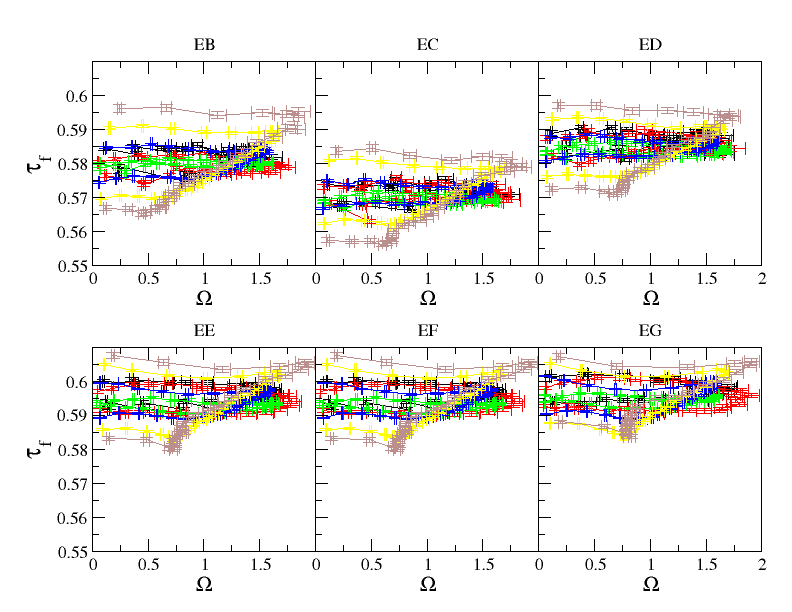}
\caption{Average friction vs average velocity  for different drive velocities (see Fig.~\ref{fig6} in the main text) }
\label{Fig6SeveralDrives}
\end{center}
\end{figure}

\begin{figure}[]
\begin{center}
\includegraphics[width=\linewidth]{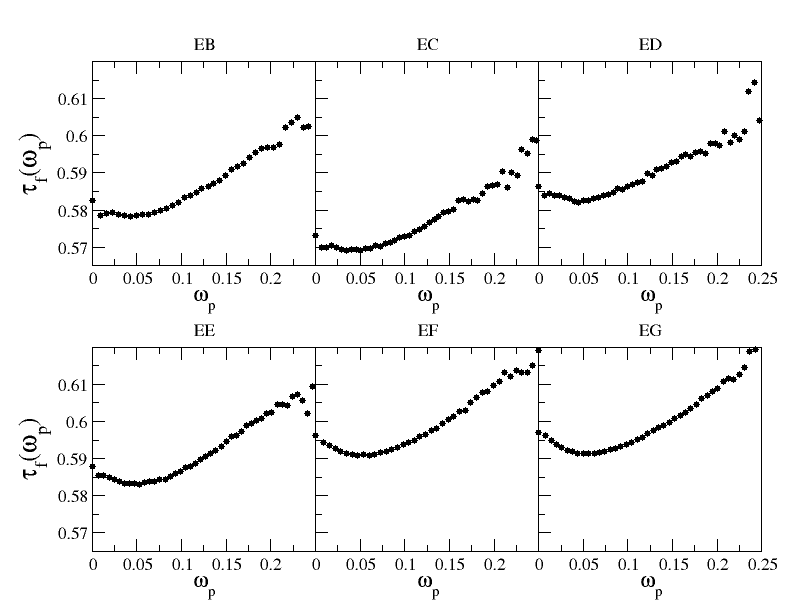}
\caption{Conditional average friction vs instantaneous plate velocity  for different drive velocities (see Fig.~\ref{fig7} in the main text) }
\label{Fig7SeveralDrives}
\end{center}
\end{figure}

The main text presents results from the series $(EA)$. The results from the other  datasets,  with the different drives reported  in  Table~\ref{tablediffdrives}, display similar behaviors and are shown 
 in Figs.~\ref{Fig2SeveralDrives}-\ref{Fig7SeveralDrives} of this Appendix, to be compared with the correspondig Figs.~\ref{fig2}-\ref{fig7} in the main text. Analogous results were  obtained adopting different sampling frequencies and threshold values.

\begin{table}
\begin{center}
  \begin{tabular}{ccc}
    duration & \# of avalanches  \\
    \hline
    0.309 $\le  T < 0.489 $& 929\\
    0.489 $\le  T < 0.722 $& 866\\
    0.772 $\le  T < 1.219 $& 987\\
    1.219 $\le  T < 1.925 $& 1694\\
    1.925 $\le  T < 3.04 $& 1380\\
    3.04 $\le  T < 4.8 $& 158\\
    \end{tabular}
\caption{\label{tableclassesEA} Classes of avalanche duration adopted for the analysis, and the resulting  number of avalanches for the  data set (EA) discussed  in the main text.}
\end{center}
\end{table}

\bibliography{sheargrbibv2}
\bibliographystyle{apsrev4-1}

\end{document}